\documentclass[aps,prl,twocolumn,floats,amsmath,preprintnumbers,nofootinbib,superscriptaddress,amssymb,10pt]{revtex4-2}
\usepackage{graphicx}
\usepackage{dcolumn}% Align table columns on decimal point
\usepackage{bm,graphicx,hyperref,subfigure}
\usepackage{float,amsfonts,graphicx}
\usepackage{color}
\usepackage{tikz}
\usetikzlibrary{decorations.pathmorphing}
\usepackage{adjustbox}
\usetikzlibrary{decorations.pathreplacing}
\usetikzlibrary{arrows.meta}
\usepackage{orcidlink}
\usepackage{ulem}
\usepackage{slashed}

\usepackage[utf8]{inputenc}
\usepackage{comment}

\usepackage{placeins}
\usepackage{xcolor}

\usepackage{cleveref} 
\crefname{figure}{Figure}{Figures}
\crefname{table}{Table}{Tables}
\crefname{section}{section}{sections} % subsection and lower inherit
\crefname{appendix}{appendix}{appendices}
\crefname{equation}{Eq.}{Eqs.}
\Crefname{figure}{Figure}{Figures}
\Crefname{table}{Table}{Tables}
\Crefname{section}{Section}{Sections} % subsection and lower inherit
\Crefname{appendix}{Appendix}{Appendices}
\Crefname{equation}{Eq.}{Eqs.}
\AddToHook{cmd/appendix/before}{\crefalias{section}{appendix}}
\relax

\newcommand{\eq}[1]{\begin{align}\label{#1}}
\newcommand{\en}{\end{align}}
\newcommand{\eqar}[1]{\begin{align}\label{#1}}
\newcommand{\enar}{\end{align}}

\newcommand{\remove}[1]{ }

\renewcommand{\vec}[1]{\bm{#1}}

\begin{document}

\setlength\abovedisplayskip{10pt}
\setlength\belowdisplayskip{10pt}

\setlength{\parskip}{14pt}
\setlength{\parindent}{0pt}

\newcommand{\Romatre}{Dipartimento di Matematica e Fisica, Universit\`a  Roma Tre and INFN, Sezione di Roma Tre,\\ Via della Vasca Navale 84, I-00146 Rome, Italy}
\newcommand{\RomatreINFN}{Istituto Nazionale di Fisica Nucleare, Sezione di Roma Tre,\\ Via della Vasca Navale 84, I-00146 Rome, Italy}
\newcommand{\Romadue}{Dipartimento di Fisica and INFN, Universit\`a di Roma ``Tor Vergata",\\ Via della Ricerca Scientifica 1, I-00133 Roma, Italy}
\newcommand{\LaSapienza}{Physics Department and INFN Sezione di Roma La Sapienza,\\ Piazzale Aldo Moro 5, 00185 Roma, Italy}
\newcommand{\soton}{Department of Physics and Astronomy, University of Southampton,\\ Southampton SO17 1BJ, UK}

\title{Rare kaon decays $K^- \to \ell^- \bar{\nu}_\ell \ell'^{+} \ell'^{-}$: Standard Model predictions from lattice QCD}

\author{R.\,Di\,Palma}\affiliation{\Romatre}
\author{R.\,Frezzotti}\affiliation{\Romadue} 
\author{G.\,Gagliardi}\affiliation{\RomatreINFN}
\author{V.\,Lubicz}\affiliation{\Romatre} 
\author{G.\,Martinelli}\affiliation{\LaSapienza}
\author{C.T.\,Sachrajda}\affiliation{\soton}
\author{F.\,Sanfilippo}\affiliation{\RomatreINFN}
\author{S.\,Simula}\affiliation{\RomatreINFN}
\author{N.\,Tantalo}\affiliation{\Romadue}

\begin{abstract}
Weak decays of charged kaons with an additional lepton-antilepton pair, $K^- \to \ell^- \bar{\nu}_\ell \ell'^{+} \ell'^{-}$ ($K_{\ell2\ell'}$), are suppressed at order $O(G_{F}^{2}\alpha_{\rm em}^{2})$ in the Standard Model (SM) and provide sensitive probes of its flavour structure, as well as independent determinations of the Cabibbo angle $|V_{us}|$. In this Letter we present the  SM predictions for all four channels with $\ell,\ell' =e,\mu$, based on the first complete lattice QCD calculation of the structure-dependent form factors reported in a companion paper~\cite{DiPalma:lattice}. Using the PDG value~\cite{ParticleDataGroup:2024cfk} $|V_{us}|^{\rm PDG}=0.22431(85)$,
we obtain branching fractions with controlled uncertainties and precisions ranging from $2\%$ to $7\%$, depending on the channel. For the three modes with published measurements, our results agree with experiment. For the $K_{\mu2\mu}$ mode, for which no published experimental result is available, we compare our prediction with the preliminary  NA62 result, finding agreement at the $1.4\sigma$ level. Conversely, the measured decay rates can be used together with our results to extract $|V_{us}|$ from these modes.  A weighted average over the two most precise channels, $K_{\mu2e}$ and $K_{\mu2\mu}$, yields $|V_{us}|=0.2283(42)$, corresponding to a $1.8\%$ determination. These results pave the way for using  $K_{\ell2\ell'}$ decays
as precision probes of the SM.
\end{abstract}

\maketitle

\section{Introduction}
The kaon sector has played a central role in establishing the flavour structure of the Standard Model (SM). Notable examples include the suppression of flavour-changing neutral currents in $K^0$--$\bar K^0$ mixing, which led to the Glashow--Iliopoulos--Maiani mechanism~\cite{Glashow:1970gm} and anticipated the existence of the charm quark, as well as the observation of the CP-violating decay $K_{L}\to\pi\pi$~\cite{PhysRevLett.13.138}, which established the presence of a complex phase in the Cabibbo--Kobayashi--Maskawa (CKM) matrix.

Kaon physics continues to provide a sensitive laboratory for testing the SM and probing possible new physics (NP) effects~\cite{Aebischer:2025mwl}. Charged-current weak decays, in particular, play a central role in determining the Cabibbo angle $|V_{us}|$~\cite{Cabibbo:1963yz}, with the most precise extractions obtained from $K_{\ell3}$ and $K_{\ell2}$ decays~\cite{FlavourLatticeAveragingGroupFLAG:2024oxs, ParticleDataGroup:2024cfk}.\footnote{A crucial role in this respect is played by lattice-QCD determination of the required isospin-breaking-corrections~\cite{Giusti:2017dwk}.} Comparisons among existing determinations of $|V_{us}|$ reveal a persistent $2-3\sigma$ tension, often referred to as the Cabibbo-angle anomaly (CAA), between extractions from $K_{\ell2}$, $K_{\ell3}$, superallowed nuclear $\beta$ decays through CKM unitarity~\cite{Hardy:2020qwl}, and $\Delta S=1$ inclusive $\tau$ decays~\cite{ExtendedTwistedMass:2024myu}. Independent determinations of $|V_{us}|$ are therefore very important to clarify these differences.

Radiative leptonic kaon decays, $K_{\ell2\gamma}$, provide additional avenues for determining $|V_{us}|$. The lattice QCD framework developed in Ref.~\cite{Desiderio:2020oej} has enabled first-principles studies of processes with an explicit photon in the final state, and $K^- \to \ell^- \bar{\nu}_\ell \gamma$ decays have been recently investigated within lattice QCD in Refs.~\cite{DiPalma:2025iud,Christ:2025ufc}. These decays depend on structure-dependent (SD) non-perturbative contributions associated with photon emission by the quarks, which lift the helicity suppression present in $K_{\ell2}$ decays and provide sensitivity to a broader class of non-standard interactions.

A natural extension of these studies is provided by the four-body leptonic decays
\begin{equation}
K^- \to \ell^- \bar{\nu}_\ell \ell'^{+}\ell'^{-},
\qquad \ell,\ell' = e,\mu,
\end{equation}
in which the photon is emitted off shell and subsequently converts into a lepton--antilepton pair. We denote these modes by $K_{\ell2\ell'}$. Although these decays occur at tree level in the SM, their rates are suppressed by two powers of the electromagnetic coupling and scale as $O(G_F^2 \alpha_{\mathrm{em}}^2)$. With branching fractions of order $10^{-8}$, they provide sensitive probes of the SM. The $K_{\ell2\ell'}$ decay amplitudes contain a helicity-suppressed point-like term proportional to the kaon decay constant $f_{K}$, as well as a SD component parameterized by four non-perturbative form factors. These decays allow for independent determinations of $|V_{us}|$ and provide additional tests of lepton-flavour universality. They have also been discussed in the context of NP scenarios involving a new mediator $X$ produced in $K^- \to \ell^- \bar{\nu}_{\ell} X$ with subsequent decay $X \to \ell^{\prime +}\ell^{\prime -}$, or in cascade processes with heavy neutral leptons~\cite{Goudzovski:2022vbt,Krnjaic:2019rsv}. 

Experimentally, three of the four leptonic channels have been measured by the BNL E865 experiment~\cite{Poblaguev:2002ug,Ma:2005iv}, while improved measurements, including the remaining mode $K_{\mu2\mu}$, are being pursued by the NA62 Collaboration~\cite{NA62:2026amj}. These developments call for precise SM predictions for both total rates and invariant-mass distributions.

These decays have traditionally been studied within Chiral perturbation theory (ChPT), which provides predictions for the SD form factors at $O(p^4)$~\cite{Bijnens:1992en}. However, the assessment of higher-order effects is limited by the poor knowledge of higher-order low-energy constants. The only first-principles method providing systematically improvable predictions is lattice QCD. Exploratory lattice studies of $K_{\ell2\ell'}$ decays have been performed in Refs.~\cite{Tuo:2021ewr,Gagliardi:2022szw} at unphysical quark masses (with $m_{K} < 2m_{\pi}$) and on a single gauge ensemble; consequently, an assessment of systematic uncertainties was not possible.

In the companion paper~\cite{DiPalma:lattice}, we present the first complete lattice QCD calculation of the full set of form factors at physical quark masses, including controlled continuum and infinite-volume extrapolations. For physical quark masses, the presence of on-shell $\pi\pi$ and $\pi\pi\pi$ intermediate states requires solving the non-trivial problem of analytic continuation to Euclidean spacetime for dilepton invariant masses above $2m_\pi$. We resolve this problem using the spectral function reconstruction (SFR) method of Ref.~\cite{Frezzotti:2023nun} in combination with the Hansen--Lupo--Tantalo (HLT) method~\cite{Hansen:2019idp}.\footnote{We refer to  Ref.~\cite{DiPalma:lattice} for technical details.}

In this Letter, we discuss the phenomenological implications of these results. Using the lattice determination of the form factors, we provide the first complete SM predictions for decay rates and differential distributions in all four channels with controlled uncertainties, enabling direct comparison with existing and forthcoming experimental measurements.

\section{Methods}

We write the decay amplitude for
\begin{equation}
K^-(p)\to \ell^-(p_\ell)\,\bar{\nu}_\ell(p_\nu)\,\ell'^+(p_+)\,\ell'^-(p_-)
\end{equation}
at leading order in the electroweak interactions as
\begin{equation}
\mathcal M(P_f)=\mathcal M_{\rm FSR}(P_f)+\mathcal M_{\rm had}(P_f),
\label{eq:Mdecomp}
\end{equation}
where $P_f\equiv\{p_\ell,p_\nu,p_+,p_-\}$ denotes the set of final-state momenta.

%------------------------------------------------------------
\begin{figure*}[t]
\centering
\includegraphics[width=1.9\columnwidth]{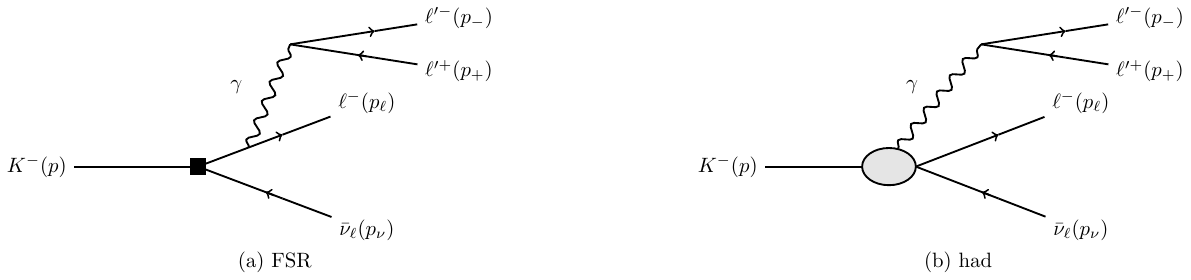}
\caption{Leading-order diagrams contributing to
$K^- \to \ell^- \bar\nu_\ell \ell'^+\ell'^-$:
(a) final-state radiation from the charged lepton line;
(b) emission of the virtual photon from the kaon.
For $\ell=\ell'$ the exchange contribution must be included (see Eq.~\ref{eq:exchange}). The black box in the left panel corresponds to the insertion of the four-fermion weak operator of the Fermi effective theory.}
\label{fig:diag}
\end{figure*}
%------------------------------------------------------------

The final-state-radiation (FSR) contribution, in which the virtual photon is emitted from the charged lepton line (Fig.~\ref{fig:diag} left), is given by
\begin{equation}
\label{eq:M_FSR}
\mathcal M_{\rm FSR}(P_{f})
=
\frac{G_F}{\sqrt2}\,V_{us}\,f_K\,\frac{e^{2}}{k^{2}}\,
L^\mu(P_{f})\,
\bar u(p_-)\gamma_\mu v(p_+)~,
\end{equation}
where $L^\mu(P_{f})$ is the leptonic kernel describing virtual-photon emission from the charged lepton line, $k=p_+ + p_-$, and $f_{K}$ is the kaon decay constant. We use $G_F=1.1663788(6)\times10^{-5}\,\mathrm{GeV}^{-2}$, and $\alpha_{\mathrm{em}}^{-1}=4\pi/e^{2} = 137.036$~\cite{ParticleDataGroup:2024cfk}.

The hadronic contribution, corresponding to photon emission from the kaon (Fig.~\ref{fig:diag} right), reads
\begin{equation}
\mathcal M_{\rm had}(P_{f})
=
-\frac{G_F}{\sqrt2}\,V_{us}\,\frac{e^2}{k^2}\,
l_\nu(P_{f})\,
\bar u(p_-)\gamma_\mu v(p_+)\,
H_W^{\mu\nu}(k,p)~,
\label{eq:Mhad}
\end{equation}
where $l_\nu(P_{f})$ denotes the charged weak leptonic current. The hadronic tensor is defined as
\begin{equation}
H_W^{\mu\nu}(k,p)
=
\int d^4x\,e^{ik\cdot x}
\langle 0|T\{j_{\rm em}^\mu(x)\,j_W^\nu(0)\}|K^-(p)\rangle~,
\label{eq:HWdef}
\end{equation}
where $j^{\mu}_{\rm em}$ is the hadronic part of the electromagnetic current and $j^{\nu}_{W}= \bar{u}\gamma^{\nu}(1-\gamma^{5})s$ is the charged weak hadronic current. Following Ref.~\cite{Desiderio:2020oej} it can be decomposed as
\begin{equation}
\label{eq:H_deco}
H_W^{\mu\nu}(k,p)=H_{\rm pt}^{\mu\nu}(k,p)+H_{\rm SD}^{\mu\nu}(k,p)~,
\end{equation}
where the point-like term $H^{\mu\nu}_{\rm pt}$, which is proportional to $f_{K}$, corresponds to the hadronic tensor obtained by treating the kaon as a point-like particle in scalar QED, while the SD part $H^{\mu\nu}_{\rm SD}$, which encodes the non-trivial hadronic dynamics, is parametrized by four SD form factors $F_V$, $F_A$, $H_1$, and $H_2$, which depend on two Lorentz invariants. We parameterize them in terms of
$x_k=\sqrt{k^2}/m_K$ and $y_k=2|\vec k|/m_K$, where $\boldsymbol{k}$ is evaluated in the kaon rest frame. 

For channels with identical negatively charged leptons in the final state, $\ell=\ell'$, one must include the exchange contribution via
\begin{align}
\label{eq:exchange}
\mathcal M(P_{f})\to \mathcal M(P_{f})-\mathcal M(P'_f)~,
\end{align}
where  $P'_f\equiv\{p_-,p_\nu,p_+,p_\ell\}$. The explicit expressions for $L^\mu(P_{f})$, $l^\mu(P_{f})$, as well as the explicit decomposition of the hadronic tensor $H_{W}^{\mu\nu}(p,k)$ are given in the Supplemental Material.

The differential decay rate is then given by
\begin{equation}
d\Gamma
=
\frac{1}{2m_K}\,
\overline{|\mathcal M|^2}\,
d\Phi_4,
\label{eq:master_rate}
\end{equation}
where $m_{K}$ is the charged-kaon mass, $d\Phi_4$ is the invariant four-body phase-space measure, and the bar denotes the lepton-spin sum. We parameterize the four-body phase space in terms of five Lorentz-invariant variables, following Refs.~\cite{Tuo:2021ewr,Gagliardi:2022szw}. The explicit expressions for the phase-space measure, kinematic variables employed,  and integration limits  are given in the Supplemental Material.

The non-perturbative inputs entering the decay rates are $m_{K}$, $f_{K}$, and the four SD form factors $F_{V}(x_k,y_{k}), F_{A}(x_k,y_k), H_{1}(x_k,y_k)$ and $H_{2}(x_k, y_k)$. For the charged-kaon mass we use the experimental value $m_{K}=493.677(13)~{\rm MeV}$ from PDG~\cite{ParticleDataGroup:2024cfk}, while for $f_{K}$ we use the determination by the Extended Twisted Mass Collaboration (ETMC), $f_{K}=156.4(6)~{\rm MeV}$~\cite{ExtendedTwistedMass:2021qui}. The remaining nonperturbative inputs are provided by the lattice QCD determination of the four SD form factors presented in the companion paper~\cite{DiPalma:lattice}. That calculation is performed on ETMC $N_f=2+1+1$ ensembles at the physical point, including an estimate of quark-disconnected contributions and controlled continuum and infinite-volume extrapolations based on three lattice spacings ($0.056~{\rm fm}\lesssim a\lesssim 0.080~{\rm fm}$) and three spatial volumes ($3.8~{\rm fm}\lesssim L\lesssim 7.7~{\rm fm}$).\footnote{The calculation of Ref.~\cite{ExtendedTwistedMass:2021qui} has been performed employing a definition of isospin-symmetric QCD that, within the uncertainties on the input parameters used, coincides with that used in~\cite{DiPalma:lattice}.}  

The hadronic tensor in Eq.~(\ref{eq:HWdef}) is obtained in~\cite{DiPalma:lattice} from suitable time-dependent Euclidean three-point correlation functions computed on the lattice (Eqs. (25)-(26) of~\cite{DiPalma:lattice}). For $x_k<2m_\pi/m_K$, the hadronic tensor is reconstructed, following the strategy already applied in previous studies of radiative processes~\cite{DiPalma:2025iud,Gagliardi:2022szw,Desiderio:2020oej}, directly from the time integration of these correlators (Sec. IV-A of~\cite{DiPalma:lattice}). For $x_k\geq 2m_\pi/m_K$, instead, the presence of intermediate states with energies below the external kinematics prevents a direct reconstruction of the Minkowski hadronic tensor in Eq.~(\ref{eq:HWdef}) from Euclidean-time integration (Sec. III of Ref.~\cite{DiPalma:lattice}). 

To circumvent this problem we employ the SFR method of Ref.~\cite{Frezzotti:2023nun} in combination with the HLT method of Ref.~\cite{Hansen:2019idp}. In this kinematic region the form factors become complex, and we evaluate both their real and imaginary parts.\footnote{The imaginary parts are numerically smaller than the real parts over the phase space relevant for the decays.}

\begin{figure}[t]
\centering
 \includegraphics[width=0.99\columnwidth]{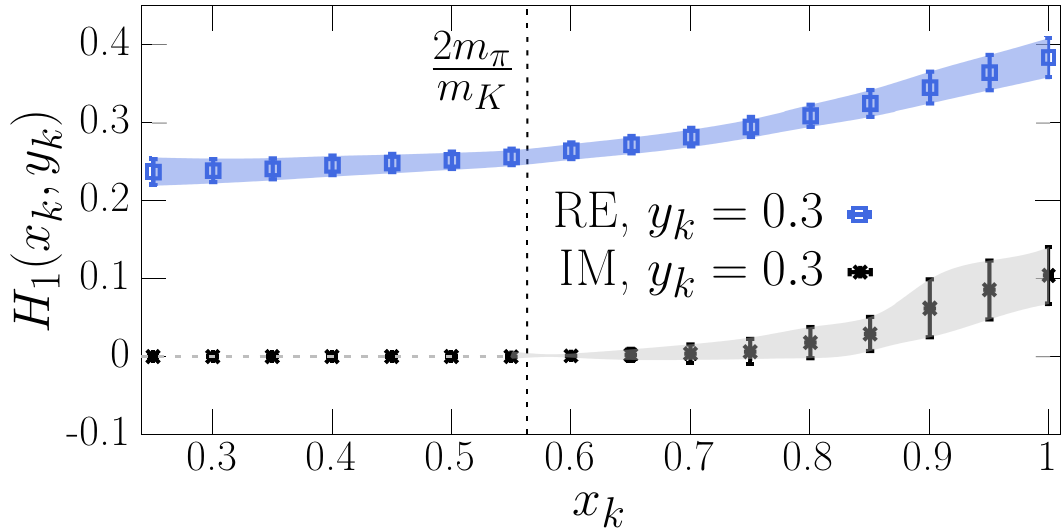}
\caption{Lattice data for the real (open squares) and imaginary (filled squares) part of $H_{1}$ vs $x_{k}$ for $y_{k}=0.3$. Bands correspond to the interpolating functions used for the calculation of the decay rates. }
\label{fig:H1}
\end{figure}

The form factors are computed at a set of discrete equally-spaced kinematic points in the $(x_k,y_k)$ plane, covering the region $x_k\gtrsim m_\pi/m_K$. The region $x_k<m_\pi/m_K$ is not included, as it contributes only to the $e^+e^-$ channels and in these cases is experimentally excluded due to the large background from semileptonic $K^- \to \pi^0 \ell^- \bar{\nu}_\ell$ decays~\cite{Poblaguev:2002ug}. To evaluate the form factors over the full kinematic region, we interpolate the lattice data using linear fits in $y_k$ at fixed $x_k$, exploiting the mild dependence of the form factors on $y_k$, followed by cubic splines in $x_k$ at fixed $y_k$.  This procedure yields an accurate representation of the form factors, with uncertainties that closely follow those of the lattice data. As an illustration, in Fig.~\ref{fig:H1} we show the lattice results for the form factor $H_1$, which provides the largest contribution to the decay rate, together with the corresponding interpolating functions. A convenient polynomial parameterization of the form factors is given in Eqs.~(182) and~(183) and Table IV of \cite{DiPalma:lattice}.

The four-body phase-space integrals are evaluated with the GSL implementation of VEGAS~\cite{Lepage:1977sw}. The correlation among the lattice data for the SD form factors is correctly propagated at all steps through a jackknife analysis. 

%------------------------------------------------------------
%------------------------------------------------------------

\section{Results}
The total decay rates $\Gamma(K_{\ell2\ell'})$
are obtained by numerically integrating the differential decay rates over the
four-body phase space. Our final results for the corresponding branching fractions are
obtained by dividing the decay rates by the total charged kaon width
$\Gamma(K^{\pm})=5.3167(86)\times 10^{-17}~{\rm GeV}$~\cite{ParticleDataGroup:2024cfk}
and by using the PDG value\footnote{The PDG value for $|V_{us}|$ is obtained combining the $K_{\ell2}$ and $K_{\ell3}$ determinations.}
$|V_{us}|^{\rm PDG}=0.22431(85)$. For $\mu^+\mu^-$ final states we obtain
\begin{align}
\label{eq:br_mumu}
{\rm Br}(K_{e2\mu}) &= 1.021(67)_{\rm SD}(2)_{K}(8)_{V_{us}}[68]\times 10^{-8}~, \nonumber \\[6pt]
{\rm Br}(K_{\mu2\mu}) &= 1.226(46)_{\rm SD}(6)_{K}(9)_{V_{us}}[48]\times 10^{-8}~,
\end{align}
where the first source of uncertainty is due to the SD form factors, the second
to $f_{K}$, $m_{K}$ and $\Gamma(K^{-})$, the third to the input value of $|V_{us}|$, and the last is the total error. 

For $e^{+}e^{-}$ final states, we impose a lower cut on the invariant mass of
the $e^{+}e^{-}$ pair, $\sqrt{k^{2}}\geq 140~{\rm MeV}$ (or $x_{k}\geq 0.284$),
matching the experimental conditions of Ref.~\cite{Poblaguev:2002ug}. For $\ell'=\ell=e$, the same cut is imposed on both $x_{k}$ and on the
exchanged variable
$x'_{k}=\sqrt{(p_{\ell}+p_{+})^{2}}/m_{K}$. We obtain
\begin{align}
\label{eq:br_ee}
{\rm Br}_{x_{k}^{(')} \geq 0.284}(K_{e2e})
&= 2.89(19)_{\rm SD}(0)_{K}(2)_{V_{us}}[19]\times10^{-8}~, \nonumber \\[6pt]
{\rm Br}_{x_{k} \geq 0.284}(K_{\mu2e})
&= 7.84(15)_{\rm SD}(5)_{K}(6)_{V_{us}}[17]\times10^{-8}~,
\end{align}
where the error breakdown is analogous to that in Eq.~(\ref{eq:br_mumu}).

The branching fractions are predicted with $2\%-7\%$ precision, depending on the channel. The uncertainty is dominated by the SD form factors. The point-like contribution alone, obtained by setting $H^{\mu\nu}_{\rm SD}(k,p)=0$, is helicity-suppressed (see Supplemental Material) and contributes about $50\%$ of the total rate in $K_{\mu2e}$, about $30\%$ in $K_{\mu2\mu}$, and negligibly in channels with an electron antineutrino. 

\begin{figure*}[t]
\centering
\includegraphics[width=1.99\columnwidth]{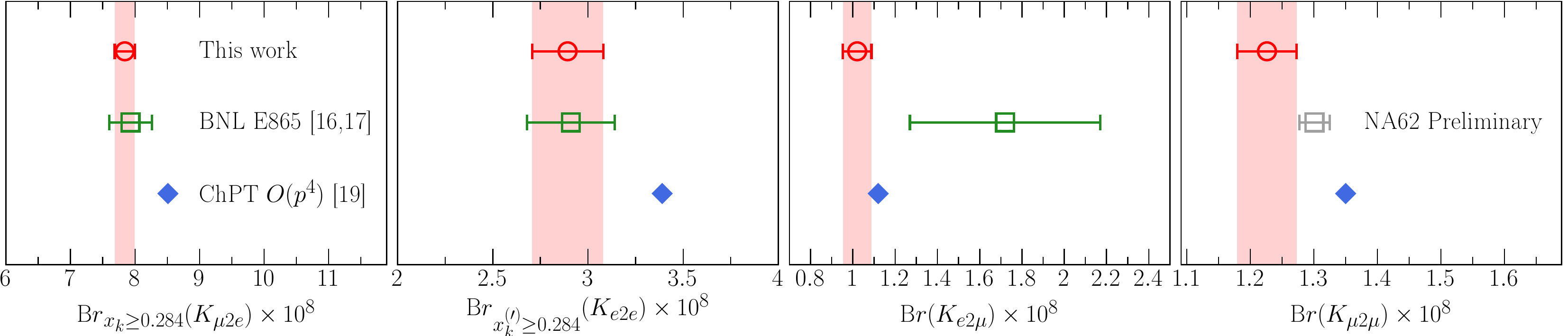}
\caption{Comparison of the SM predictions for the $K_{\ell2\ell'}$ branching fractions with experimental
measurements from BNL E865 and the preliminary NA62 result. The ChPT
predictions at $O(p^{4})$ from Ref.~\cite{Bijnens:1992en} are also shown.}
\label{fig:comp_exp}
\end{figure*}

Our results for all channels but the purely muonic one can be compared with
the BNL E865 measurements~\cite{Poblaguev:2002ug,Ma:2005iv}.\footnote{For decays with electrons in the final state, radiative corrections of $O(\alpha_{\rm em}^{3}\log[m_K/m_e])$ may be significant~\cite{Christ:2025ufc}. These corrections, which depend on experimental acceptances and efficiencies, are estimated by the experimental collaborations~\cite{Poblaguev:2002ug,Ma:2005iv,NA62:2026amj} and are not included in our $O(\alpha_{\rm em}^{2})$ theoretical predictions.} In the purely muonic channel the only
published bound,
${\rm Br}(K_{\mu2\mu})\leq 4.1\times 10^{-7}$
from Ref.~\cite{PhysRevLett.63.2177}, is far above our prediction, but we can nevertheless compare it with 
the preliminary NA62 result~\cite{NA62:2026amj}. The comparisons are
shown in Fig.~\ref{fig:comp_exp}, together with the ChPT predictions at
$O(p^{4})$ from Ref.~\cite{Bijnens:1992en}. As the figure shows, we find
excellent agreement with experiment for the $e^{+}e^{-}$ channels.
The same conclusion holds when comparing with the BNL E865 results obtained
with the alternative cuts $\sqrt{k^2}\geq145~{\rm MeV}$ and
$\sqrt{k^2}\geq150~{\rm MeV}$ (see Supplemental Material).
For $K_{e2\mu}$, our result is below the experimental one, but the difference is not significant, amounting to a $1.5\sigma$ effect. In the purely muonic channel, $K_{\mu2\mu}$, the comparison
with the preliminary NA62 result indicates an agreement at the level of $1.4\sigma$. The $O(p^{4})$ ChPT predictions from Ref.~\cite{Bijnens:1992en} are systematically larger than our results by about $8\%\text{--}17\%$, depending on the channel, with the largest difference observed for $K_{\mu2e}$.
Although the relative difference is at the $8\%$ level in this case, the higher precision of our results makes the difference correspond to about a $4\sigma$ effect when compared to our uncertainties.
%------------------------------------------------------------
%------------------------------------------------------------

\begin{figure}[t]
\centering
\includegraphics[width=0.9\columnwidth]{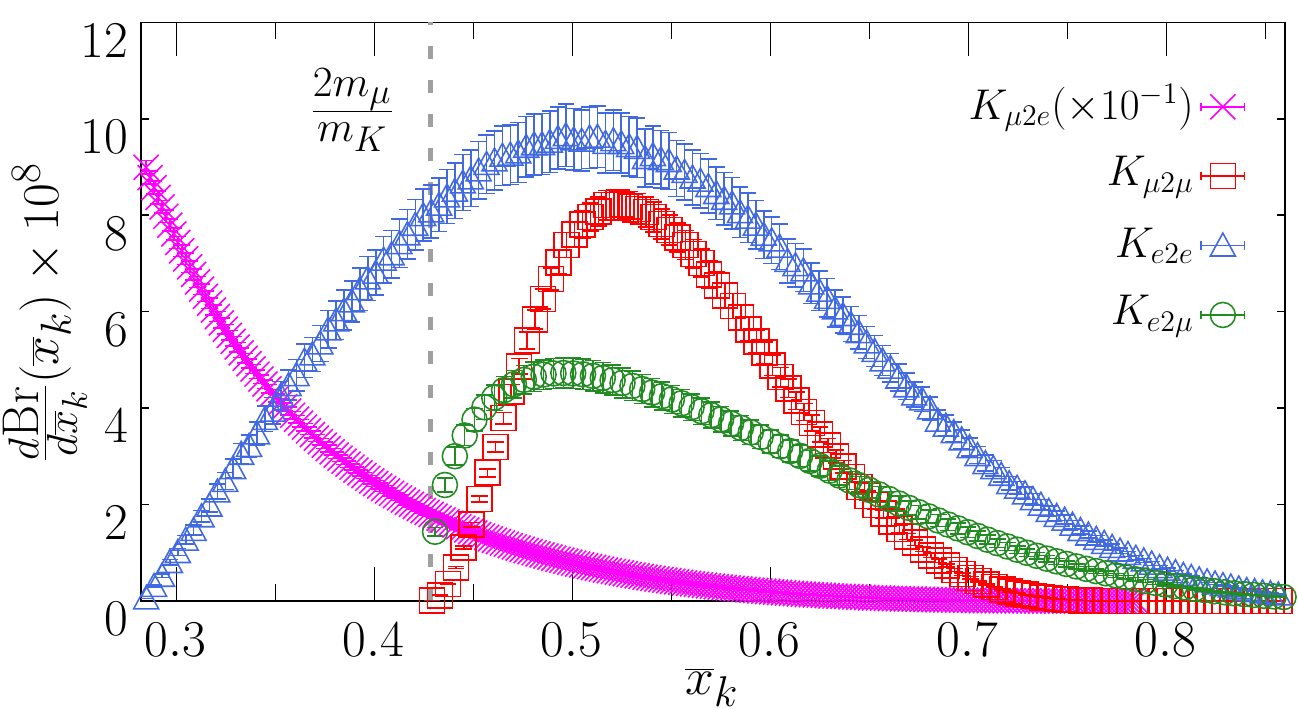}
\caption{Differential branching fractions for 
$K_{\ell2\ell'}$ 
as defined in Eq.~(\ref{eq:diff_branching}). 
The data corresponding to $K_{\mu2e}$ have been divided by  10 for better visualization.}
\label{fig:differential_results}
\end{figure}

In Fig.~\ref{fig:differential_results} we show the invariant mass distributions of the $\ell'^{+}\ell'^{-}$ pair, defined as
\begin{align}
\label{eq:diff_branching}
\frac{d{\rm Br}}{d\overline{x}_{k}}(\overline{x}_{k})=
\frac{1}{\Gamma(K^{-})}\frac{d}{d\overline{x}_{k}}
\int_{x_{k}^{(')}\leq \overline{x}_{k}} d\Gamma~,
\end{align}
where in the case of equal leptons, $\ell=\ell'$, the integral in Eq.~(\ref{eq:diff_branching}) is restricted to the region $x_{k}, x'_{k} \leq \overline{x}_{k}$, and for $K_{e2e}$ a further cut $x_{k}^{(')} \geq 0.284$ is implied.
These quantities are experimentally accessible and provide information beyond
the total branching fractions, offering direct
sensitivity to the kinematic dependence of the SD form factors.
They also enhance the sensitivity to possible NP effects, which could
manifest as localized distortions or resonant features in the invariant-mass
spectrum and might be diluted in integrated observables. The definition in Eq.~(\ref{eq:diff_branching}) is also straightforward to implement experimentally in the equal-lepton case, $\ell=\ell'$: for each event, one measures both $x_k$ and $x'_k$ and fills the histogram with $\max(x_k,x'_k)$. Differential branching fractions at selected $\overline{x}_{k}$ points  are provided in tabular form in the Supplemental Material. 

%------------------------------------------------------------
%------------------------------------------------------------

The comparisons discussed so far use the PDG value for the Cabibbo angle
$|V_{us}|$. Interesting quantities that are independent of both $|V_{us}|$ and $\Gamma(K^-)$
 are the lepton-flavour-universality (LFU)
ratios of branching fractions. With four channels, six
ratios can be formed. We denote them by
\begin{align}
\label{eq:LFU}
R_{\ell_{2}\ell'_{2}}^{\ell_{1}\ell'_{1}} \equiv 
\frac{ {\rm Br}(K_{\ell_{1}2\ell'_{1}}) }
{ {\rm Br}(K_{\ell_{2}2\ell'_{2}}) }~,
\end{align}
where, as before, for $e^{+}e^{-}$ final-states the branching fractions are evaluated with the cut $x_{k}^{(')} \geq 0.284$.
The LFU ratios benefit from partial cancellation of correlated uncertainties in the SD form factors between numerator and denominator.  Our results for the six LFU ratios are reported in
Table~\ref{tab:LFU} together with the experimental values.

\begin{table}[t]
\begin{ruledtabular}
\begin{tabular}{cccccc}
& th & exp & & th & exp \\
$R_{ee}^{e\mu}$ & $0.3528(73)$ & $0.59(16)$ & $R^{e\mu}_{\mu e}$ & $0.1302(65)$ & $0.217(57)$ \\
$R_{ee}^{\mu e}$ & $2.71(13)$ & $2.72(24)$ & $R^{e\mu}_{\mu\mu}$ & $0.832(28)$ & $1.32(35)$ \\
$R_{ee}^{\mu\mu}$ & $0.424(17)$ & $0.447(36)$ & $R_{\mu e}^{\mu\mu}$ & $0.1564(33)$ & $0.1641(75)$ \\
\end{tabular}
\caption{SM predictions (th) and experimental values (exp) for the six independent LFU ratios
$R_{\ell_{2}\ell'_{2}}^{\ell_{1}\ell'_{1}}$ defined in
Eq.~(\ref{eq:LFU}). \label{tab:LFU}}
\end{ruledtabular}
\end{table}

\begin{figure}[t]
\centering
\includegraphics[width=0.90\columnwidth]{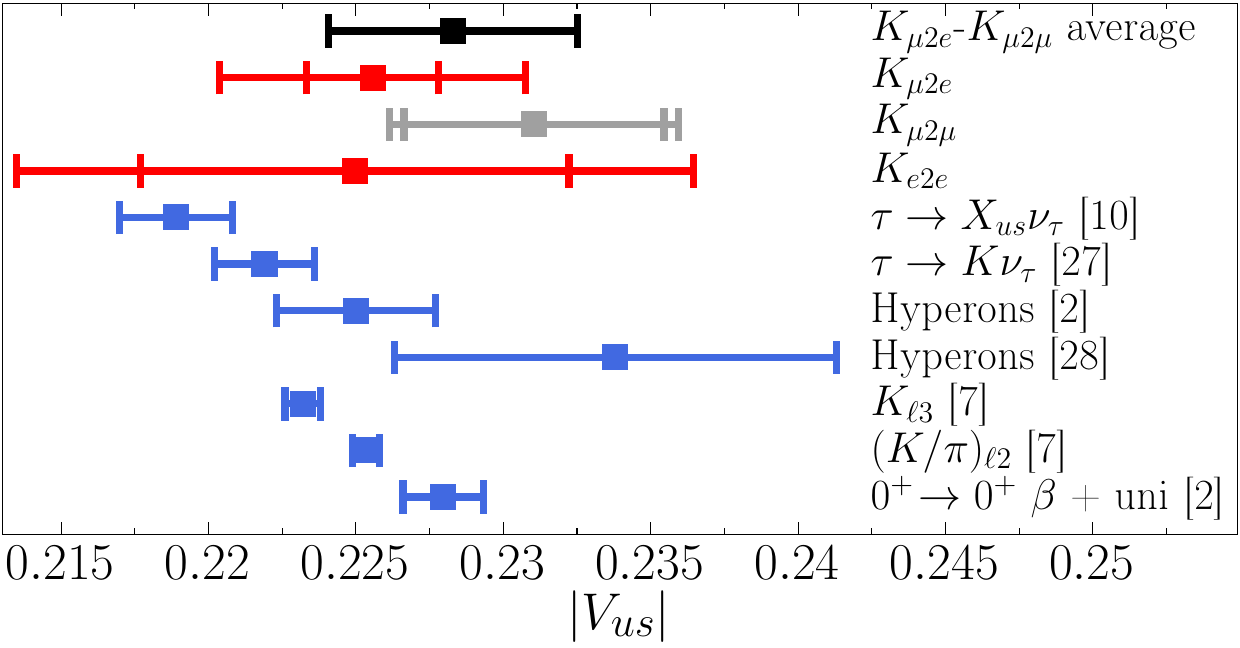}
\caption{$|V_{us}|$ values obtained in this work (excluding the noisy $K_{e2\mu}$ result), together with existing determinations. Inner error bars denote theory uncertainties.}
\label{fig:comp_Vus}
\end{figure}

Finally, the comparison with the experimental results can be turned into an independent determination of $|V_{us}|$. We define ${\rm Br}_{0}$ by factoring out the overall $|V_{us}|^{2}$ dependence from the theoretical branching fraction. The corresponding
determination of $|V_{us}|$ then follows from (the square root of) the ratio between the measured
branching fraction and ${\rm Br}_{0}$. For the three channels measured by the
BNL E865 experiment we obtain
\begin{align}
|V_{us}|_{\mu2e} &= 0.2256(22)_{\rm th}(47)_{\rm exp}[52]~, \nonumber \\[6pt]
|V_{us}|_{e2e} &= 0.225(7)_{\rm th}(9)_{\rm exp}[11]~, \nonumber \\[6pt]
|V_{us}|_{e2\mu} &= 0.291(10)_{\rm th}(38)_{\rm exp}[39]~,
\end{align}
where the first and second source of uncertainties correspond to theory and experimental errors, respectively, and the last is their combination.
The extraction from $K_{\mu2e}$
reaches a precision of about $2.3\%$ and is currently dominated by the
experimental uncertainty, so an improved measurement by the NA62 Collaboration
would be very welcome. Similarly, if the NA62 preliminary result in the purely
muonic channel is confirmed, the corresponding determination,
$|V_{us}|_{\mu2\mu}=0.2310(44)_{\rm th}(21)_{\rm exp}[49]$, would also reach a comparable precision. A summary plot containing
the three most precise determinations of $|V_{us}|$ from this work, together with existing
determinations~\cite{ParticleDataGroup:2024cfk,FlavourLatticeAveragingGroupFLAG:2024oxs,ExtendedTwistedMass:2024myu,HFLAV:2022esi,Bacchio:2025auj}, is shown in Fig.~\ref{fig:comp_Vus}. Given the preliminary
nature of the experimental result for the purely muonic channel, we represent the
corresponding data point in gray. The black data point, corresponding to $|V_{us}| =0.2283(42)$, is the weighted average of the $|V_{us}|$ values from $K_{\mu2e}$ and $K_{\mu2\mu}$, taking into account their correlation.

%------------------------------------------------------------
%------------------------------------------------------------

\section{Conclusions}

In this Letter, together with the companion paper~\cite{DiPalma:lattice},
we have presented the first complete SM predictions for
$K^{-}\to \ell^{-}\bar{\nu}_{\ell}\ell'^{+}\ell'^{-}$ decay rates based on a
first-principles lattice QCD determination of the required SD contributions.
The calculation combines standard methods to extract the relevant form
factors from Euclidean lattice correlators with novel spectral density reconstruction
techniques, allowing us to treat the kinematic region above the two-pion threshold. This work constitutes
the first complete determination of an exclusive electroweak hadron decay employing spectral reconstruction methods.

Comparison with existing measurements shows good agreement in all channels. We have also
shown that determinations of $|V_{us}|$ from these modes already reach a
precision at the $1.8\%$ level after combining the results from the $K_{\mu2e}$ and $K_{\mu2\mu}$ modes, despite the small branching fractions of
$O(10^{-8})$. The current uncertainties, depending on the channel considered, are dominated by experimental errors and/or by the statistical precision of the SD form factors. Since both can be improved in the future, there is room for improving the precision of the $|V_{us}|$ determination from these modes.

The strategy employed here is general. In particular, spectral density
reconstruction techniques allow one to treat electroweak amplitudes in the presence of an arbitrary number of multi-particle
intermediate states below the external kinematics. This opens up the possibility of applying the same
approach to analogous decays of heavier pseudoscalar mesons, such as
$D_{(s)}$ and $B$ mesons, which we plan to investigate in future work.

\section{Acknowledgements}
We thank the ETMC for a very fruitful collaboration. VL, RF, GG, FS, SS and NT have been supported by the Italian Ministry
of University and Research (MUR) and the European
Union (EU) – Next Generation EU, Mission 4, Component 1, PRIN 2022, CUP F53D23001480006. 
FS is supported by ICSC – Centro Nazionale di Ricerca in High Performance Computing, Big Data and Quantum Computing, funded by European Union - Next Generation EU and by Italian  Ministry of University and Research (MUR) project FIS\_00001556. We acknowledge support from the LQCD123, ENP, and SPIF Scientific Initiatives of
the Italian Nuclear Physics Institute (INFN). CTS is partially supported by STFC(UK) consolidated grant ST/X000583/1.

\bibliography{biblio}

\appendix
\section{SUPPLEMENTAL MATERIAL}

\subsection{Details of the decay amplitude}
In this section we provide explicit expressions for the quantities entering the final-state-radiation contribution, $\mathcal{M}_{\rm FSR}(P_{f})$, and the hadronic contribution, $\mathcal{M}_{\rm had}(P_{f})$, to the total amplitude $\mathcal{M}(P_{f})$ introduced in Eq.~(\ref{eq:Mdecomp}).

The weak leptonic current $l^\nu(P_{f})$ appearing in Eq.~(\ref{eq:Mhad}) is
\begin{align}
l^\nu(P_{f})=\bar u(p_\ell)\gamma^\nu(1-\gamma^5)v(p_\nu)~,
\end{align}
where $\bar u$ and $v$ denote the spinors of the charged lepton $\ell^-$ and antineutrino $\bar\nu_\ell$, respectively. The leptonic kernel $L^\mu(P_{f})$ is defined as
\begin{equation}
\label{eq:L_def}
L^\mu(P_{f})=l^\mu(P_{f})+
m_\ell\,\bar u(p_\ell)
\frac{\slashed{k}\gamma^\mu+2p_\ell^\mu}
{m_\ell^2-(k+p_\ell)^2}
(1-\gamma^5)v(p_\nu)~,
\end{equation}
where $m_\ell$ is the mass of the charged-lepton $\ell^{-}$.

The hadronic tensor $H_W^{\mu\nu}(k,p)$ is decomposed into a point-like term $H_{\rm pt}^{\mu\nu}(k,p)$ and a structure-dependent part $H_{\rm SD}^{\mu\nu}(k,p)$ (Eq.~(\ref{eq:H_deco})).
Their explicit expressions are
\begin{align}
\label{eq:H_pt}
H_{\rm pt}^{\mu\nu}(k,p)
=
f_K\left[
g^{\mu\nu}
+
\frac{(2p-k)^\mu (p-k)^\nu}{ m_{K}^{2} - (p-k)^{2}}
\right]
\end{align}
and
\begin{align}
\label{eq:H_SD}
H_{\rm SD}^{\mu\nu}(k,p)
={}&
-i\frac{F_V}{m_K}\varepsilon^{\mu\nu\alpha\beta}k_\alpha p_\beta
+\frac{H_1}{m_K}\left(k^2g^{\mu\nu}-k^\mu k^\nu\right)
\nonumber \\[8pt]
&+\frac{F_A}{m_K}\left[(p\cdot k-k^2)g^{\mu\nu}-(p-k)^\mu k^\nu\right]
\nonumber \\[8pt]
&+\frac{H_2}{m_K}
\frac{(p\cdot k-k^2)k^\mu-k^2(p-k)^\mu}{(p-k)^2-m_K^2}
(p-k)^\nu~,
\end{align}
where $F_V$, $H_1$, $F_A$, and $H_2$ are the four non-perturbative SD form factors, parameterized in terms of the two Lorentz scalars $x_{k}$ and $y_{k}$, determined in the companion paper~\cite{DiPalma:lattice}. We use the convention $\varepsilon^{0123}=1$.

The point-like term saturates the electromagnetic Ward--Takahashi identity satisfied by the hadronic tensor,
\begin{align}
k_\mu H_{\rm pt}^{\mu\nu}(k,p)
=
i\langle 0|j_W^\nu|K^-(p)\rangle
=
f_K p^\nu~,
\end{align}
while the SD contribution is transverse, namely $k_\mu H_{\rm SD}^{\mu\nu}(k,p)=0$. The point-like contribution to the amplitude $\mathcal{M}(P_{f})$, obtained by setting $H_{\rm SD}^{\mu\nu}(k,p)=0$, is helicity suppressed. This can be seen as follows. First, the $g^{\mu\nu}$ term in $H_{\rm pt}^{\mu\nu}(k,p)$ in Eq.~(\ref{eq:H_pt}) exactly cancels the term arising from the $l^\mu(P_{f})$ contribution to $L^\mu(P_{f})$ in Eq.~(\ref{eq:L_def}). Second, the remaining term in $H_{\rm pt}^{\mu\nu}(k,p)$ contains the factor
$
(p-k)^\nu=(p_\ell+p_\nu)^\nu$,
which, when contracted with the weak leptonic current $l_\nu(P_{f})$, gives a contribution that vanishes in the massless limit $m_\ell\to0$ by virtue of the equations of motion,
\begin{align}
(p_\ell+p_\nu)^\nu l_\nu(P_{f})
=
m_\ell\,\bar u(p_\ell)(1-\gamma^5)v(p_\nu)~.
\end{align}
For the same reason, the contribution of the form factor $H_2$ to the amplitude is also helicity suppressed, since it contains the same factor $(p-k)^\nu$.

\subsection{Parameterization of the four-body phase space}

In this section we provide details of the parameterization of the four-body phase space $d\Phi_{4}$ introduced in Eq.~(\ref{eq:master_rate}). We start from the standard Lorentz-invariant definition
\begin{align}
d\Phi_{4} &= \mathcal{S}\,(2\pi)^{4}\,
\delta^{(4)}(p-p_{-}-p_{+}-p_{\ell}-p_{\nu})
 \times \nonumber \\[6pt]
&\times \prod_{r=\{\ell,\nu,+,-\}}\frac{d^{3}\boldsymbol{p}_{r}}{(2\pi)^{3}\, 2p^{0}_{r}} ,
\end{align}
where $\mathcal{S}$ is a symmetry factor equal to $1/2$ for identical charged leptons in the final state ($\ell=\ell'$) and $\mathcal{S}=1$ otherwise. The four-momenta are denoted by $p_{r}^{\mu}=(p^{0}_{r},\boldsymbol{p}_{r})$ for $r=\{\ell,\nu,+,-\}$.

The four-body phase space can be equivalently expressed in terms of five Lorentz-invariant variables. Following Ref.~\cite{Tuo:2021ewr}, it can be written as
\begin{align}
d\Phi_{4}= \frac{\mathcal{S}\,2\omega\, m_{K}^{4} y_{k}}{2^{14}\pi^{6}}
\, dx_{k}\,dx_{q}\,dy_{12}\,dy_{34}\,d\phi ,
\end{align}
where $k=p_{+}+p_{-}$ and
\begin{align}
x_{k}=\frac{\sqrt{k^{2}}}{m_{K}}, \qquad
x_{q}=\frac{\sqrt{(p-k)^{2}}}{m_{K}} ,
\end{align}
are the normalized invariant masses of the $\ell'^{+}\ell'^{-}$ and $\ell\bar{\nu}_{\ell}$ systems, respectively. The dimensionless variable $y_{k}$ introduced in the main text is related to $x_{k}$ and $x_{q}$ by
\begin{align}
y_{k} = \sqrt{(1-x_{k}^{2}-x_{q}^{2})^{2} - 4x_{k}^{2}x_{q}^{2}}
      = \sqrt{\lambda(1,x_{k}^{2},x_{q}^{2})} ,
\end{align}
where $\lambda(x,y,z)$ is the Källén function, while $\omega=2x_{k}x_{q}$.

The remaining variables are defined as
\begin{align}
y_{12} &= \frac{2}{m_{K}^{2}y_{k}}(p_{+}-p_{-})\cdot(p_{\ell}+p_{\nu}), \nonumber\\[6pt]
y_{34} &= \frac{2}{m_{K}^{2}y_{k}}\,k\cdot
\left[\left(1+\frac{r_{\ell}^{2}}{x_{q}^{2}}\right)p_{\nu}
+\left(1-\frac{r_{\ell}^{2}}{x_{q}^{2}}\right)p_{\ell}\right], \nonumber\\[6pt]
\sin\phi &= 
\frac{16}{\omega m_{K}^{4} y_{k}}
\frac{\varepsilon_{\mu\nu\rho\sigma}
p_{+}^{\mu}p_{-}^{\nu}p_{\nu}^{\rho}p_{\ell}^{\sigma}}
{\sqrt{(\lambda_{12}^{2}-y_{12}^{2})
(\lambda_{34}^{2}-y_{34}^{2})}},
\end{align}
where $r_{\ell}=m_{\ell}/m_{K}$ and $r_{\ell'}=m_{\ell'}/m_{K}$, with $m_{\ell'}$ being the mass of the lepton $\ell'^{\pm}$.\footnote{Note that with respect to Ref.~\cite{Gagliardi:2022szw} we redefined $r_{\ell^{(')}} \to  r_{\ell^{(')}}^{2}$.}  The quantities
\begin{align}
\lambda_{12}=\sqrt{1-\frac{4r_{\ell'}^{2}}{x_{k}^{2}}}, \qquad
\lambda_{34}=1-\frac{r_{\ell}^{2}}{x_{q}^{2}},
\end{align}
determine the allowed ranges of $y_{12}$ and $y_{34}$. 

The integration domain is given by
\begin{align}
2r_{\ell'}\leq x_{k}\leq 1-r_{\ell}, \qquad
r_{\ell}\leq x_{q}\leq 1-x_{k},
\end{align}
together with
\begin{align}
-\lambda_{12}\leq y_{12}\leq \lambda_{12}, \qquad
-\lambda_{34}\leq y_{34}\leq \lambda_{34}, 
\end{align}
and $\phi\in[0,2\pi]$.

To evaluate the decay rate within a given region of the phase space we perform a Monte Carlo integration over the five variables $x_{k},x_{q},y_{12},y_{34},\phi$ using the GSL implementation of the VEGAS algorithm. A C++ implementation of the integrator, together with the code used to evaluate the unpolarized squared matrix element $\overline{|\mathcal{M}|^{2}}$ as a function of the non-perturbative inputs and of the five phase-space variables, is publicly available at
\href{https://github.com/GGagliardi/KlllnuIntegrator}{https://github.com/GGagliardi/KlllnuIntegrator}.
\vspace{2cm}

\section{Comparison with BNL E865 for alternative dilepton-mass cuts}

In addition to the cut $\sqrt{k^{2}}\geq 140~{\rm MeV}$ discussed in the main
text, Ref.~\cite{Poblaguev:2002ug} reports results for the $e^{+}e^{-}$
channels with $\sqrt{k^{2}}\geq145~{\rm MeV}$ and
$\sqrt{k^{2}}\geq150~{\rm MeV}$. 

For the $K_{e2e}$ channel, which contains two identical negatively charged
electrons, the cut is applied to both dilepton invariant masses, i.e. we require
$\sqrt{k^{2}},\sqrt{k'^{2}}\geq\sqrt{k_{\rm min}^2}$, where
$k=p_{+}+p_{-}$ and $k'=p_{\ell}+p_{+}$. We have repeated the
phase-space integrations with these cuts using the same setup described in the
main text.

The resulting comparison between our SM predictions and the BNL E865
measurements is shown in Table~\ref{tab:ee_cuts_supp}.  The agreement is
excellent for both channels and for both cuts.
\begin{center}
\begin{table}[h]
\begin{ruledtabular}
\begin{tabular}{ccc}
$\sqrt{k_{\rm min}^2}$ [MeV] & This work & E865 \\
\hline
\multicolumn{3}{c}{${\rm Br}(K_{\mu2e})\times 10^{8}$} \\[2pt]
\cline{2-3}
$145$ & $6.97(14)_{\rm SD}(4)_{K}(5)_{V_{us}}[15]$ & $7.06(31)$ \\
$150$ & $6.20(13)_{\rm SD}(4)_{K}(5)_{V_{us}}[14]$ & $6.28(27)$ \\
\multicolumn{3}{c}{${\rm Br}(K_{e2e})\times 10^{8}$} \\[2pt]
\cline{2-3}
$145$ & $2.68(17)_{\rm SD}(0)_{K}(2)_{V_{us}}[17]$ & $2.70(22)$ \\
$150$ & $2.48(16)_{\rm SD}(0)_{K}(2)_{V_{us}}[16]$ & $2.48(20)$ \\
\end{tabular}
\end{ruledtabular}
\caption{Comparison between our SM predictions and the BNL E865 measurements
for the $e^{+}e^{-}$ channels with alternative cuts on the dilepton invariant
mass. The breakdown of the theoretical uncertainties is analogous to that
discussed in the main text.}
\label{tab:ee_cuts_supp}
\end{table}
\end{center}

\section{Numerical values for the differential decay widths}
In this section we provide in Table~\ref{tab:diff_branching} the differential
branching fractions defined in Eq.~(\ref{eq:diff_branching}) for selected values
of $\overline{x}_{k}$. Differential branching fractions evaluated at values of
$\overline{x}_{k}$ other than those reported in
Table~\ref{tab:diff_branching}, as well as differential branching fractions
integrated over selected bins, are available from the authors upon request.

\onecolumngrid

\begin{center}
\begin{table}[h]
\begin{ruledtabular}
\begin{tabular}{ccccc}
& \multicolumn{4}{c}{$d\mathrm{Br}/d\overline{x}_{k} \times 10^{8}$} \\
\cline{2-5}
$\overline{x}_{k}$
& $K_{e2e}$
& $K_{e2\mu}$
& $K_{\mu2e}$
& $K_{\mu2\mu}$ \\[6pt]
$0.35$ & $4.13(29)$ & -- & $42.37(84)$ & -- \\
$0.40$ & $6.81(47)$ & -- & $24.52(59)$ & -- \\
$0.45$ & $8.87(59)$ & $3.73(23)$ & $14.19(42)$ & $1.807(55)$ \\
$0.50$ & $9.57(64)$ & $4.72(32)$ & $7.93(29)$ & $7.58(27)$ \\
$0.55$ & $9.05(59)$ & $4.18(29)$ & $4.12(18)$ & $7.66(31)$ \\
$0.60$ & $7.48(48)$ & $3.31(24)$ & $1.919(94)$ & $4.92(21)$ \\
$0.65$ & $5.25(33)$ & $2.30(17)$ & $0.690(37)$ & $2.096(96)$ \\
$0.70$ & $3.19(19)$ & $1.42(11)$ & $0.1520(87)$ & $0.489(23)$ \\
$0.75$ & $1.612(89)$ & $0.764(57)$ & $0.00808(50)$ & $0.0254(12)$ \\
$0.80$ & $0.648(35)$ & $0.339(28)$ & -- & -- \\
\end{tabular}
\end{ruledtabular}
\caption{Differential branching fractions
$d\mathrm{Br}/d\overline{x}_{k}$ defined in Eq.~(\ref{eq:diff_branching}),
evaluated at selected values of $\overline{x}_{k}$. Missing entries (denoted by ``--'')
correspond to kinematically forbidden regions.}
\label{tab:diff_branching}
\end{table}
\end{center}

\twocolumngrid

\end{document}